\documentclass[12pt,epsf]{article}
\pdfoutput=1

\usepackage{mathtools,amsthm}

\usepackage{amsfonts}
\usepackage{amsmath}
\usepackage[colorlinks=true, urlcolor=dblue,linktocpage=true,linkcolor=dblue,citecolor=purple]{hyperref}
\usepackage{ulem}
\usepackage{tipa}
\usepackage{graphics,graphicx,xcolor}
\usepackage{setspace}
\usepackage{subcaption}
\usepackage{mathrsfs,amssymb}

\colorlet{dblue}{blue!70!black}

\newcommand{\nn}{\nonumber}

\newcommand{\lb}{\left(}
\newcommand{\rb}{\right)}

\newcommand\be{\begin{equation}}
\newcommand\ba{\begin{eqnarray}}
\newcommand\ee{\end{equation}}
\newcommand\ea{\end{eqnarray}}

\newcommand{\arxivold}[1]
  {\href{http://arxiv.org/abs/#1}{#1}}
\newcommand{\arxiv}[1]
  {\href{http://arxiv.org/abs/#1}{arXiv:#1}}

\newcommand{\set}[1]{\left\{#1\right\}}
\newcommand{\tr}[1]{\text{Tr }#1}
\newcommand{\partialtrace}[2]{\text{Tr}_#1 \; #2}

\begin{document}

\begin{spacing}{1.3}
\begin{titlepage}

\begin{center}
{\Large \bf Holographic Entropy Cone Measures}

\vspace*{6mm}

Ning Bao,$^*$ Samuel Blitz,$^\dagger$ and Bogdan Stoica$^\ddagger$

\vspace{5mm}

\textit{
$^*$ Institute of Quantum Information and Matter, California Institute of Technology, Pasadena, CA 91125, USA
}\\ 

\vspace{2mm}

\textit{
$^*$ Walter Burke Institute for Theoretical Physics,\\ California Institute of Technology, 452-48, Pasadena, CA 91125, USA
}\\ 

\vspace{2mm}

\textit{
$^\dagger$ Department of Physics,\\ University of California, Davis, CA 95616, USA
}\\ 

\vspace{2mm}

\textit{
$^\ddagger$ Martin A. Fisher School of Physics,\\ Brandeis University, Waltham, MA 02453, USA
}\\

\vspace{2mm}

\textit{
$^\ddagger$ Department of Physics, Brown University, Providence RI 02912, USA
}\\

\vspace{5mm}

{\tt bstoica@brandeis.edu, ningbao@its.caltech.edu, shblitz@ucdavis.edu}

\vspace*{1cm}
\end{center}
\begin{abstract}

We investigate numerically several proxy measures for the number of states contained within the holographic entropy cone, compared to the number contained within the quantum entropy cone, for states with $3$ and $4$ parties. We find an interesting tension: while measures focused on calculating the volume ratios between the two cones indicate that the quantum cone is much larger than the holographic one, measures based on the generation of random states and then calculating the entropies thereof imply that almost all such randomly generated states are also contained within the holographic entropy cone. Also interestingly, the volume measures strongly indicate a difference in the number of states within the quantum or stabiliser cones versus the number in the holographic cone, which is not reproduced by the other classes of measures. We comment on the difference between the two classes of measures, and why each may be preferable.

\end{abstract}
\vspace{1cm}

\begin{flushleft}{\qquad Brown-HET-708, BRX-TH-6312, CALT-TH 2017-001}\end{flushleft}

\end{titlepage}
\end{spacing}

\vskip 1cm

\setcounter{tocdepth}{2}


\vfill\eject

\begin{spacing}{1.3}

\section{Entanglement Entropy and Holography}
In a quantum theory (conformal field, quantum mechanical, etc.), one can define the Von Neumann entropy as
\begin{align*}
S = -\tr{\rho \ln{\rho}},
\end{align*}
where $\rho$ is the density matrix of some quantum state. What makes this definition interesting is the ability to perform a partial trace over part of a Hilbert space, to obtain a mixed state for which $S$ measures the correlations between the two parts of the Hilbert space. For example, if $\mathcal{H} = \mathcal{H}_A \otimes \mathcal{H}_B$, we can obtain the reduced density matrix $\rho_A$ by tracing over $\mathcal{H}_B$, i.e. 
\begin{align*}
\rho_A = \partialtrace{B}{\rho}.
\end{align*}

In the AdS/CFT correspondence \cite{Maldacena:1997re,Gubser:1998bc,Witten:1998qj}, the entanglement entropies of regions of holographic states have special geometric interpretations, according to the Ryu-Takayanagi formula \cite{RT}: on a surface of time-reflection symmetry, the entanglement entropy between a region $R$ of the boundary CFT and its complement $R^C$ is proportional to the area of the minimal codimension-2 surface in the bulk homologous to $R$,
\begin{align*}
S = \frac{A_{min}}{4G}.
\end{align*}
 Using this relationship, one can prove several inequalities about entanglement entropy using geometric constructions (see, for example, \cite{SSA}). These proof techniques should and do successfully reproduce all inequalities that hold for all quantum systems, namely Araki-Lieb, subadditivity, strong subadditivity, and weak monotonicity. However, using these same techniques it is possible to prove other inequalities that are known to hold, not for all quantum states, but for states for which the Ryu-Takayanagi formula applies. The first of these ``holographic entanglement entropy inequalities'' to be found was the monogamy of mutual information (MMI) \cite{MMI}, and more have been discovered since - including an infinite family thereof \cite{HEC}.

Because these are known inequalities that do not in general hold for any quantum (or CFT) state, but do hold for any state with a holographic dual, a good litmus test to determine whether a state can have a holographic dual is to compute the entanglement entropies for various regions and check if they satisfy the set of inequalities known to only hold for holographic states. If the regions of the test state do not satisfy all of these inequalities, the state certainly does not have a geometric holographic dual. We note, however, that this test is a necessary but not sufficient condition for holography.

It is interesting, then, to try to gauge (an upper bound on) what ``fraction'' of all states can have holographic duals. The following sections will detail our attempts to answer this question.

\section{Cones and Measures}

When considering the entropic inequalities satisfied by a certain class of states, it is useful to introduce the notion of an entropy cone. For a partition of the Hilbert space $\mathcal{H} = \mathcal{H}_{A_1} \otimes \dots  \mathcal{H}_{A_n}$ given by some parties $A_1,\dots,A_n$, one can compute the entanglement entropy $S\lb A_{i_1} \dots A_{i_k} \rb$ of the union of parties $A_{i_1},\dots,A_{i_k}$ with their complement, and construct an entropy ray $R$ (a $(2^n-1)$-dimensional vector),
\be
R = \lb S(A_1), \dots, S(A_1\dots A_n) \rb \nn.
\ee
The $n$-party entropy cone is the union of the entropy rays for all allowed partitions of the Hilbert space by $n$ parties, and for all states in the class under consideration. In this paper, we will be interested in the entropy cone for generic quantum states (QEC), for stabiliser states (SEC), and for holographic states (HEC). One technical detail to note is that entropy vectors for $n$ parties are identified under permutations and purifications by the permutation group $S_{n+1}$.

In order to determine the fraction of all states that can have holographic duals, one first needs to decide on the measure with which one considers the states. In the simplest case of a quantum state comprised of several qubits, one can generate $n$-qubit states with random coefficients for each element in the computation basis, compute all $2^n-1$ entanglement entropies and check whether the resulting entropy ray falls inside the HEC. This particular measure, which we call the ``uniform state measure,'' is one way to quantify an upper bound on the fraction of quantum states that have holographic duals.

Another measure that can be used to quantify an upper bound on the fraction of holographic states is to randomly generate the entropy vectors themselves. Obviously, one must then check that the entropy vectors have a quantum state associated with them (i.e. they lie within the quantum entropy cone). However, this measure is independent of the theory in which the states lie - all quantum (CFT or otherwise) states must have entropy vectors that lie in the quantum entropy cone, so the specifics of theory cannot obscure or skew any data obtained. A downside of this measure, however, is that the quantum states in a given theory are not in a one-to-one correspondence with an entropy vector - any given entropy vector (which satisfies the quantum entropy cone) can have anywhere between a single and infinite number of states associated with it. In fact, for every point within the quantum entropy cone, there exist non-holographic states - even when that point is contained in the HEC. We call this measure the ``cone measure.''

Another measure is to enumerate a list of representatives of equivalence families of qubit states (with equivalence generated by determinant 1 SLOCC operations) - these are finite, and known up to 4-qubit systems \cite{VERSTAETE}. From these representative qubit states, one can examine which of these families of states satisfy the HEC and which do not - we call this the ``finite family measure.'' This does not directly answer the question of what fraction of quantum states can be holographic, but it may provide insight in that direction. A drawback of considering equivalence classes of states is that different states in the same family can have different entanglement entropies, as pointed out by \cite{mukund-max}, who found that only one family in the SLOCC classification is on the same side of MMI.\footnote{A different partition into families, for random stabiliser states on three parties, that classifies the entanglement structure was recently proposed in \cite{Nezami:2016zni}.} We will not consider the finite family measure in this paper.

Finally, we could also simply enumerate over all $n$-qubit states with coefficients from a finite set, and determine which of these states correspond to entropy vectors which lie in the HEC. This measure will be referred to as the ``exhaustive $n$-qubit measure.''

\section{Results}
To determine the (upper bound) on the fraction of holographic states using the uniform state measure, one can simply generate random states and check whether they satisfy the HEC. This was done previously (see \cite{mukund-max}) in the case where the HEC was simply taken to be $\text{QEC} \cap \text{MMI}$. The result of that computation was that, as the number of qubits increased, the fraction of states that satisfied MMI asymptotically approached 1 -- even with only 4 qubits, random states tended to satisfy MMI. Extending this calculation to include all of the 5-party holographic inequalities proven in \cite{HEC}, we have found that these inequalities are still not very restrictive: out of $10^3$ trial states, all satisfied these inequalities as well.

Given enough randomly generated points, the uniform state measure should have similar qualities regarding the fraction of which satisfy the HEC as those determined via the exhaustive qubit measure. Using the exhaustive 4-qubit measure with coefficients taken from the set $\set{0,1}$ (and then properly normalized), we found that
\begin{align*}
\frac{v(\text{HEC})}{v(\text{QEC})} =0.967,
\end{align*}
so the vast majority of states satisfy MMI.

Finally, using the cone measure, we found strikingly different results: in the 3-party case, we found that
\begin{align*}
\frac{v(\text{HEC})}{v(\text{QEC})} = 0.571 \pm 0.004,
\end{align*}
where we have also quoted the standard deviation, and for the 4-party case,
\begin{align*}
\frac{v(\text{HEC})}{v(\text{QEC})} > 0.076\pm 0.005.
\end{align*}
The full 4-party quantum cone is not known, so the 4-party quantum entropy cone stated above is simply an intersection of the known valid 4-party entanglement entropy inequalities. The true 4-party quantum cone will be smaller, but it can be lower bounded via the 4-party stabiliser cone (the entropy cone for stabiliser states, \cite{LindenRuskai:2013}), which it is known to contain, which in turn contains the 4-party holographic cone. For completeness, we present that here:
\begin{align*}
\frac{v(\text{HEC})}{v(\text{SEC})} = 0.078 \pm 0.005 ,
\end{align*}
where we note that in this measure the 4-party SEC and the QEC are very close,
\be
\frac{v(\text{SEC})}{v(\text{QEC})} = 0.98 \pm 0.03. \nn
\ee

Thus, the cone measure strongly differentiates between the quantum and stabiliser cones and the holographic one, whereas the state measures give the three cones to be of roughly the same ``size.''

We are limited by computational resources for higher-party states, so we could not check the relative size of the 5-party HEC. However, we expect the fractional size of the HEC relative to the QEC to continue to decrease.

\section{Discussion}
It is worth remarking that not all measures produce the same result. Indeed, while the uniform state, finite family and exhaustive $n$-qubit measure all seem to asymptote to indicating that most quantum states have entropies consistent with being holographic, the cone measure seems to indicate quite the opposite. This can be understood by noting that when generating states in Hilbert space, it is much more likely to land close to states that have entropy vectors given by the so-called star configurations (entropy vectors that can be realized holographically, with duals consisting of star graphs) \cite{mukund-max}. However, having such a discrepancy between multiple intuitively sensible measures indicates a tension as to which one is preferable.

The strengths and weaknesses of the two classes of measures are clear. The state measures have the advantage of directly relating to specific, though low qubit-number quantum states. The fact that they are associated with states with small numbers of qubits, however, is a hindrance when arguing that they should be representative of infinite dimensional CFT states, which clearly do not have small numbers of qubits. It is possible that behaviors of small $n$ qubit systems and infinite $n$ qubit systems are very different, especially from the perspective of the poorly understood notion of multipartite entanglement. 

In fact, a specific issue with the application of the uniform state measure to this point has been the assumption that the purifying system is a single qubit. In general, there is no bound on the size of the purifying system, though simulating large purifying systems will quickly prove to be computationally intractable. Despite this, however, a possible direction for future work is to fix the number of qubits of the mixed state, and to study the effect of increasing the number of qubits in the state that purifies to the mixed state. It will be interesting to check whether there is any correlation between the size of the purifying system and the position of the resulting mixed state relative to the HEC.

With the cone measure, the concerns are exactly the reverse: associating the measure to specific quantum states is difficult, and could lead to issues regarding over- or under- sampling of specific vectors. This measure, however, is agnostic as to the dimensionality of the Hilbert space being considered, and thus more naturally positioned to study infinite dimensional Hilbert spaces. Also interestingly, the cone measure seems to strongly differentiate between the holographic and stabiliser entropy cones, while indicating a relatively small difference between the stabiliser and quantum entropy cones, potentially providing an indication of the special nature of holography relative to more traditional families of quantum states. Moreover, it provides evidence that further consideration of cone volumes in this way may be able to better relate at a level beyond simply ordinal inclusion the typicality of entropies for specific classes of states.

The two classes of measures considered in this paper do not even morally have the same behavior, so one cannot conclusively use either one as of yet to declare what the ratio of potential holographic states to quantum states is. The question of which of these measures is closer to a natural counting of states is therefore an interesting one, and one that deserves more careful attention.

\section*{Acknowledgements}

We thank V.~Hubeny, H.~Maxfield and M.~Rangamani for valuable discussions. B.~Stoica would like to thank Brandeis HPC. The work of B.~Stoica is supported in part by the Simons Foundation, and by the U.S. Department of Energy under grant DE-SC-0009987. The research of N.~Bao is funded in part by the Walter Burke Institute for Theoretical Physics at Caltech, and by DOE grant DE-SC-0011632.

\end{spacing}

\end{document}